\documentclass[14pt,a4paper]{article}%
\usepackage{amsmath}
\usepackage{graphicx}
\usepackage{amsfonts}
\usepackage{amssymb}%
\begin{document}

\begin{center}
\vspace{15mm}

{\large PRIOR\ MEASURE \ FOR NONEXTENSIVE\ ENTROPY}

\bigskip

F.Brouers and O.Sotolongo-Costa\bigskip

Institute of Physics, Li\`{e}ge University and Henri Poincar\'{e} Chair of
Complex Systems, Havana University, Cuba.

fbrouers@ulg.ac.be , oscarso@ff.oc.uh.cu\bigskip

\bigskip

ABSTRACT
\end{center}

\bigskip\baselineskip0.60cm

We show that if one uses the invariant form of the Boltzmann-Shannon
continuous entropy, it is possible to obtain the generalized Pareto-Tsallis
density function, using an appropriate "prior" measure m$_{q}$(x) and a
"Boltzman constraint" which formally is equivalent to the Tsallis $q$-average
constraint on the random variable X. We derive the Tsallis prior function and
study its scaling asymptotic behavior.\ When the entropic index $q$ tends to
1$,$ m$_{q}$(x) tends to 1 for all values of x as this should be.

\bigskip


\bigskip

\bigskip

\bigskip

\section{Introduction}

Most of the probability distributions used in natural, biological, social and
economic sciences can be formally derived by maximizing the entropy with
adequate constraints ($maxS$\ principle)\cite{Kap89}.

According to the $maxS$\ principle, given some partial information about a
random variable $i.e.$\ the knowledge of related macroscopic measurable
quantities (macroscopic observables), one should choose for it the probability
distribution that is consistent with that information but has otherwise a
maximum uncertainty.\ In usual thermodynamics, the temperature is a
macroscopic observable and the distribution functions are exponentials.

Quite generally, one maximizes the Shannon-Boltzmann (S-B) entropy:
\begin{equation}
S_{B}=-\int_{a}^{b}p(x)\ln p(x)dx
\end{equation}
subject to the conditions%
\begin{equation}
\int_{a}^{b}p(x)dx=1,\int_{a}^{b}g_{i}(x)p(x)dx=<g_{i}(x)>,\text{ \ }i=1,2,..
\end{equation}
Both limits $a$and $b$ may be finite or infinite.\ The functions $g_{i}%
(x)$whose expectation value have been usually considered \cite{Kap89} as
constraints to build probability distributions are of the type
\begin{equation}
x,x^{2},x^{n},(x-<x>)^{2},\mid x\mid,\mid x-<x>\mid,\ln x,(\ln x)^{2},\ln(1\pm
x),\exp(-x)....
\end{equation}
The maximum entropy probability density function ($mepdf)$ depends on the
choice of the limits of integration $a$ and $b$ and the functions $g_{i}(x)$
whose expectation values are prescribed.

One constructs the Lagrangian
\begin{equation}
L=-\int_{a}^{b}p(x)\ln(p(x)dx
\end{equation}%
\[
{\Large -\lambda}_{0}{\Large (}\int_{a}^{b}{\Large p(x)dx-1)-}\sum
_{i}{\Large \lambda}_{i}{\Large (}\int_{a}^{b}{\Large g}_{i}%
{\Large (x)p(x)dx-<g}_{i}{\Large (x)>}%
\]
and differentiating with respect to $f(x)$, one finds easily :%

\begin{equation}
p(x)=C\exp[-\sum_{i}\lambda_{i}g_{i}(x)]
\end{equation}
The factor $C$ is a normalization constant and the Lagrange parameters
$\lambda_{i}$ are determined by using the constraints (eqs.2,3).When this
cannot be achieved simply, the parameters $\lambda_{i}$ are defined by the
constraints. Most distributions derived from the constraints given in eq.3
posses finite second moments and hence belong to the domain of attraction of
the normal distributions.\ Those which belong to the domain of attraction of
the L\.{e}vy (stable) distribution $i.e$. the Cauchy and the Pareto
distributions are obtained with a characteristic L\.{e}vy tail parameter
$\mu\geq1$ indicating that only a finite expectation value (first moment) can
be defined. In particular using the simplest constraint%
\begin{equation}
\int_{0}^{\infty}xp(x)dx=<x>
\end{equation}
one obtains the density
\begin{equation}
p(x)=(1/<x>)\exp(-x/<x>)
\end{equation}
which is the basis of Boltzmann thermostatistics if we identify $x$ with the
energy $E$ and define the temperature as $T=k<E>$ where $k$ is the Boltzmann constant.

\section{Shortcomings of the $S\max$ principle}

This traditional utilization of $maxS$ principle has two main shortcomings.

The first one is that it cannot provides distributions belonging to the basin
of attraction of Levy distributions when the Levy heavy tail index
$\mu<1.i.e.$distributions corresponding to the statistial properties of
non-averaging systems. Applied to thermostatistics this means that it cannot
be applied to nonextensive systems.

The second, pointed out repeatedly for decades by Jaynes\cite{Jay68}%
\cite{Jay03}, is that the continuous form of the Boltzman-Shannon entropy is
not invariant and should be written more correctly as
\begin{equation}
S=-\int p(x)\ln\frac{p(x)}{m(x)}dx
\end{equation}
in order to have a distribution invariant under parameter change. The function
$m(x)$ is called the measure function. It is proportional to the limiting
density of discrete points. So it is the entropy, relative to some measure,
which has to be maximized.

Under a change of variable the function $p(x$) and $m(x)$ transform in the
same way. If we maximize the entropy
\begin{equation}
\int p(x)\ln\frac{p(x)}{m(x)}dx
\end{equation}
with the constraint%
\begin{equation}
\int g(x)p(x)dx=1
\end{equation}
The Lagrange multiplier method yields for $p(x)$ the solution%
\begin{equation}
p(x)=Cm(x)e^{-\lambda g(x)}%
\end{equation}
The meaning of the measure function has been discussed at large by Jaynes
\cite{Jay68}\cite{Jay03}.\ If there is no constraint maximizing the entropy
yields $p(x)=Cm(x)$ where $C$ is a normalizing constant.\ It is the
distribution representing "complete ignorance"

\section{Tsallis entropy}

The most popular method to circumvent the first difficulty is based on a
generalization of the Boltzmann entropy known as the Tsallis
entropy\cite{Tsa88}\cite{Tsa98}.\ This method has been used with some success
to deal with thermostatistic properties of slightly chaotic and nonergodic
systems presenting memory effects and long-range interactions.

The Tsallis entropy is appropriate for system weakly chaotic power law mixing
when the Liapounov exponent is tending to zero and the relevant phase space
instead of being translational invariant (as in the Gibbs-Boltzmann case) is
scale invariant (fractal or more generally multifractal). We start from the
Tsallis nonextensive entropy%
\begin{equation}
S_{T}=-\int_{0}^{\infty}p_{q}^{q}(x)\ln_{q}p_{q}(x)dx
\end{equation}
subject to the conditions%
\begin{align}
\int_{0}^{\infty}p_{q}(x)dx  &  =1,\int_{0}^{\infty}g_{q},_{i}(x)\tilde{p}%
_{q}(x)dx=<g_{q,i}(x)>_{q},\text{ \ }i=1,2,.\ .\\
\tilde{p}_{q}(x)  &  =\frac{p_{q}^{q}(x)}{\int_{a}^{b}p_{q}^{q}(x)dx}\text{
\ \ }f_{q}(x)=\frac{\tilde{p}_{q}^{1/q}(x)}{\int_{a\ }^{b\ }\tilde{p}%
_{q}^{1/q}(x)dx}%
\end{align}
where the functions $\tilde{p}_{q}(x)(x)$ are the so-called \textquotedblright
escort\textquotedblright\ probability \cite{Tsa98}\cite{Tsa01} .\ Generalizing
what has been done with the standard S-B entropy (eqs.1-4), one can write the
constraints using the generalized $q$-exponential and $q$-logarithm functions
($q$-constraints).%
\begin{equation}
\ln_{q}x\equiv\frac{x^{1-q}-1}{1-q}\text{ \ \ \ \ \ }\exp_{q}(x)\equiv
\lbrack1+(1-q)x]^{\frac{1}{1-q}}\text{ \ \ \ with \ \ }\ln_{q}(\exp_{q}(x))=x
\end{equation}
construct the corresponding Lagrangian $\ $and differentiating with respect to
$\tilde{p}_{q}(x)$ one finds \cite{Bro04}%
\begin{equation}
p_{q}(x)=C_{q}\exp_{q}[-\sum_{i}\lambda_{q,i}g_{q,i}(x)]
\end{equation}%
\begin{equation}
\tilde{p}_{q}(x)=\tilde{C}_{q}p_{q}(x)^{q}\
\end{equation}

The factors $C_{q}$ are normalization constants and the Lagrange parameters
$\lambda_{q,i}$ are determined by using the $q$-constraints. If one chooses
the simplest constraint with $g_{q,i}(x)=x$ defining the so- called $q$-average%

\begin{equation}
\text{$\textsc{E}$}_{q}(x)=\int_{0}^{\infty}x)\tilde{p}_{q}(x)(x)dx=<x>_{q}%
\end{equation}
we get the generalized Pareto density supported on the positive half-line and
defined in the range $1<q<2.$
\begin{equation}
\tilde{p}_{q}(x)=(1/(2-q)<x>_{q})[1+\frac{\ q-1}{2-q}\ \frac{x}{<x>_{q}%
}]^{-\frac{q}{q-1}}%
\end{equation}

\begin{equation}
p_{q}(x)=(1/<x>_{q})[1+\frac{\ q-1}{2-q}\ \frac{x}{<x>_{q}}]^{-\frac{1}{q-1}}%
\end{equation}
and the corresponding Pareto-Tsallis distribution supported on the positive
half-line.%
\begin{equation}
P_{q}(x)=1-[1+\frac{\ q-1}{2-q}\ \frac{x}{<x>_{q}}]^{-\frac{q}{q-1}}%
\end{equation}
Defining the tail index $\mu=\frac{2-q}{q-1},$ the distribution $P_{q}(x)$ has
an asymptotic algebraic behavior
\begin{equation}
p_{q}(x->\infty)->(1/\mu)\ x^{-\mu-1}%
\end{equation}%
\begin{equation}
P_{q}(x->\infty)->(1/\mu)\ x^{-\mu}%
\end{equation}
The $n$-th moments E[x$^{n}]$ of the random variable of the random variable
$X$ distributed according to eq.() exists only if $n$%
$<$%
$\mu~$(hence for $1<q<\frac{2+n}{n+1}).$Therefore $P_{q}(x)$ belongs to the
domain of attraction of the completely assymmetric stable L\'{e}vy
distribution. These properties and their consequences are widely discussed in
the literature.\cite{Tsa01}\cite{Abe03}

\section{Determination of the prior for nonextensive entropy}

We can now come to the second shortcoming. The meaning of the measure function
has been discussed at large by Jaynes\cite{Jay68}\cite{Jay03}.\ In the
continuous case even before we can apply maximum entropy principle, we must
deal with the problem of complete ignorance (for instance all microstates are
equiprobable in Boltzmann-Gibbs thermodynamic).\ For "fractal" systems
ignorance means scaling invariance. It is therefore legitimate to ask what is
the scale invariant measure function or "prior" which should be used to obtain
the Generalized Pareto-Tsallis density function if we would use the correct
invariant form
\begin{equation}
S_{I}=-\int_{0}^{\infty}p_{q}(x)\ln\frac{p_{q}(x)}{m_{q}(x)}dx
\end{equation}
and use the constraint
\begin{equation}
<g(x)>=%
{\displaystyle\int}
g(x)p_{q}(x)dx
\end{equation}
Starting from the identity
\begin{equation}
S_{I}=-\int p_{q}(x)\ln\frac{p_{q}(x)}{m(x)}dx=S_{T}=\int(p_{q}^{\ }%
(x))^{q}\ln_{q}p(x)dx
\end{equation}
we derive the following relations:%
\begin{equation}
\ln p_{q}(x)-\ln m(x)=(p_{q}(x))^{q-1}\ln_{q}p_{q}(x)
\end{equation}

\[
S_{T}=S_{B}+<\ln m(x)>\
\]
or defining the "Kullback-Leibler divergence" $KL(p(x),g(x))$%

\begin{equation}
KL(p_{q}(x),m(x))=S(p_{q},m)-S_{B}\
\end{equation}
the quantity
\begin{equation}
S(p_{q},m)=-<\ln m(x)>
\end{equation}
being known as "cross-entropy".\ The cross entropy is always greater than or
equal to the entropy, this shows that the Kullback-Leibler ($KL$) divergence
is always nonnegative and furthermore $KL(p(x),p(x))$ is zero. We obtain easily%

\begin{equation}
\frac{p_{q}(x)}{m_{q}(x)}=\exp[(p(x))^{q-1}\ln_{q}p_{q}(x)]
\end{equation}
We use the definition of $\ln_{q}$ (eq.15), and obtain%

\begin{equation}
m_{q}(x)\ =p_{q}(x)\exp(\ln_{q}(\frac{1}{p_{q}(x)})
\end{equation}
If we choose for $p_{q}(x)$ the generalized Pareto-Tsallis function with
reduced variable ( $<x>_{q}=1)$%

\begin{equation}
p_{q}(x)=(1+\frac{q-1}{2-q}x)^{-\frac{1}{q-1}}%
\end{equation}
We obtain easily the following results:%

\begin{equation}
\ln_{q}(\frac{1}{p_{q}(x)})=\frac{x}{(2-q)+(q-1)x}%
\end{equation}%
\begin{equation}
m_{q}(x)=\exp_{q}(-\frac{1}{2-q}x)\exp(\frac{x}{(2-q)+(q-1)x})
\end{equation}

If we define the $q-$Laplace transform in he sense of Lenzi et al.
\cite{Len99} :
\begin{equation}
\mathcal{L}_{q}[f(t)](s)\equiv F_{q}(s)\equiv\int_{0}^{\infty}f(t)[\exp
_{q}(-t)]^{s}dt\
\end{equation}

we have
\begin{equation}
\int_{0}^{\infty}m_{q}(x)dx=F_{q}(1)=\int_{0}^{\infty}f(t)[\exp_{q}%
(-t)]dt\ \text{\ \ with }f(t)=\frac{t}{1+(q-1)t}%
\end{equation}
In term of the Levy tail index
\begin{equation}
\mu=\frac{2-q}{q-1}%
\end{equation}
we can write%

\begin{equation}
m_{q}(x)=(1+\frac{1}{\mu}x)^{-\mu-1})\exp(\frac{(\mu+1)x}{\mu+\ x})
\end{equation}
This function obeys the following asymptotic behaviors%
\begin{equation}
m_{q}(x->\infty)->\exp(1+\mu)\frac{1}{\mu}x^{-\mu-1}%
\end{equation}%
\begin{equation}
m_{q}(x->0)=1-(\mu+1)(x/\mu)^{2}+...
\end{equation}
The prior function $m_{q}(x)$ has the scaling invariant asymptotic form of a
completely asymmetric L\'{e}vy stable distribution. In Fig.1 and Fig.2 we
represent the function for two value of $q$ (1.2 and 1.8) coresponding to
values of the tail index $\mu$ ($4$ and $0.25$) and for the physical range
$1<q<2$.

\section{Maximization of the invariant entropy}

If we maximize the entropy
\begin{equation}
S_{I}=\int p_{q}(x)\ln\frac{p_{q}(x)}{m(x)}dx
\end{equation}

with the constraints%
\begin{equation}
\int_{0}^{\infty}\ p_{q}(x)dx=1\text{ \ \ \ \ \ }\int_{0}^{\infty}%
g(x)p_{q}(x)dx=1
\end{equation}
The Lagrange multiplier method yields for $p(x)$ the solution%
\begin{equation}
p_{q}(x)=Cm_{q}(x)e^{-\lambda g(x)}%
\end{equation}

We will recover the Tsallis function%
\begin{equation}
p_{q}(x)=(1+\frac{q-1}{2-q}x)^{-\frac{1}{q-1}}%
\end{equation}

if
\begin{equation}
\frac{x}{(2-q)+(q-1)x}=(1/(2-q))\frac{x}{1+(q-1)/(2-q)x}=\lambda g(x)
\end{equation}

i.e.%
\begin{equation}
\lambda=1\ ,\text{\ }(<g(x)>=1)\text{\ and }g(x)=(1/(2-q))\frac{x}%
{1+(q-1)/(2-q)x}%
\end{equation}
In that case the constraint (equation ) reads%
\begin{equation}
\int g(x)p_{q}(x)dx=(1/(2-q))\int x(1+\frac{q-1}{2-q}x)^{-\frac{q}{q-1}}dx=1
\end{equation}
which is the\ Tsallis q-average\ constraint on the variable $x$ since
\begin{equation}
\tilde{p}_{q}(x)=(1/(2-q))(1+\frac{q-1}{2-q}x)^{-\frac{q}{q-1}}%
\end{equation}
is the Tsallis escort probability (eq.19). Therefore the use of the correct
non invariant form
\begin{equation}
S_{I}=-\int_{0}^{\infty}p_{q}(x)\ln\frac{p_{q}(x)}{m_{q}(x)}dx
\end{equation}
with the "prior"%
\begin{equation}
m_{q}(x)=\exp_{q}(-\frac{1}{2-q}x)\exp(\frac{x}{(2-q)+(q-1)x})
\end{equation}
and the constraint%

\begin{equation}
\int g(x)p_{q}(x)dx=1\
\end{equation}
with%
\begin{equation}
g(x)=(1/(2-q))\frac{x}{1+(q-1)/(2-q)x}%
\end{equation}
yields the generalized Pareto density
\begin{equation}
p_{q}(x)=(1+\frac{q-1}{2-q}x)^{-\frac{1}{q-1}}%
\end{equation}
and the "Boltzmann" constraints (eq.) is equivalent to the Tsallis constraints
$<x>_{q}=1$.

\section{Conclusions}

We have shown that maximizing the invariant continuous Boltzmann-Shannon
entropy with appropriate prior measure and constraint provides the Generalized
Pareto Tsallis distribution which is the basis of the nonextensive
thermostatistic.\ This point of view opens paths for deriving superstatistics
directly from the "universal" invariant Boltzmann-Shannon entropy.

\section{Acknowledgment}

One of us is grateful to Prof.N.Kumar for useful discussions.

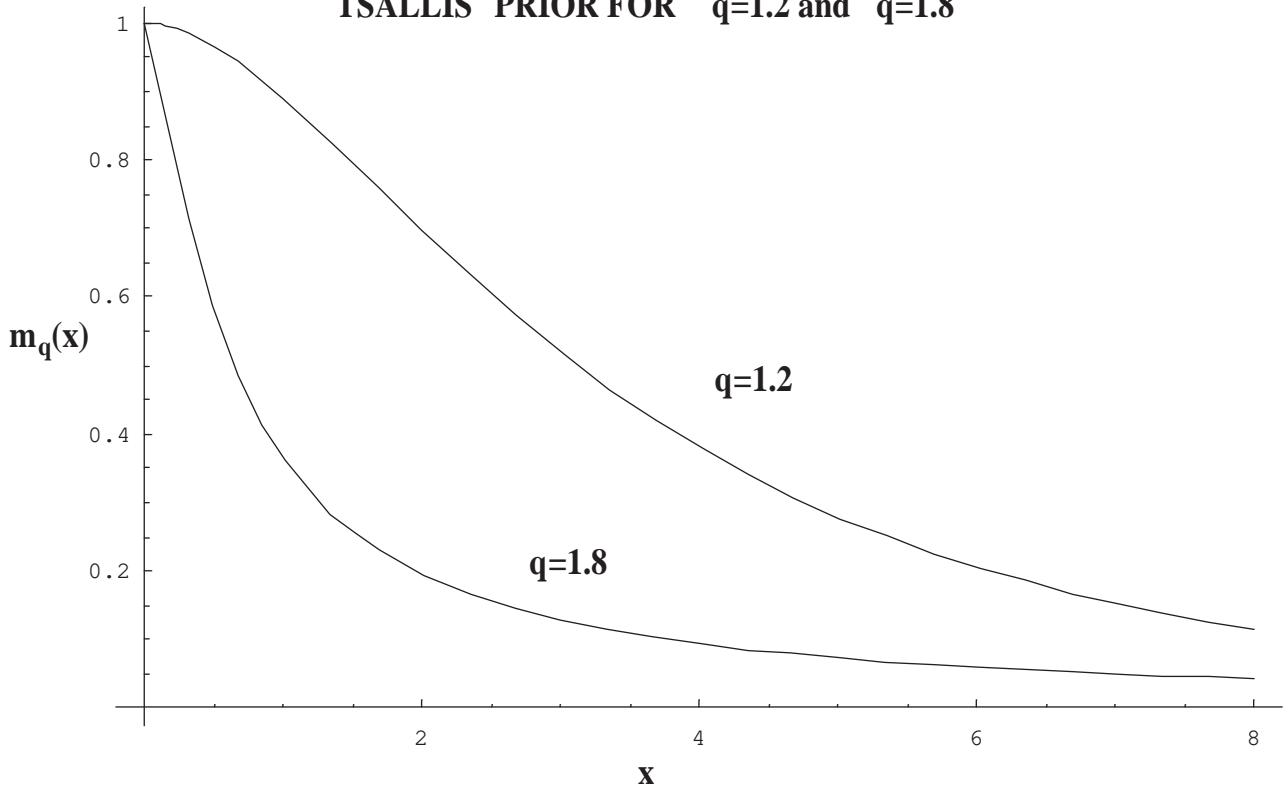
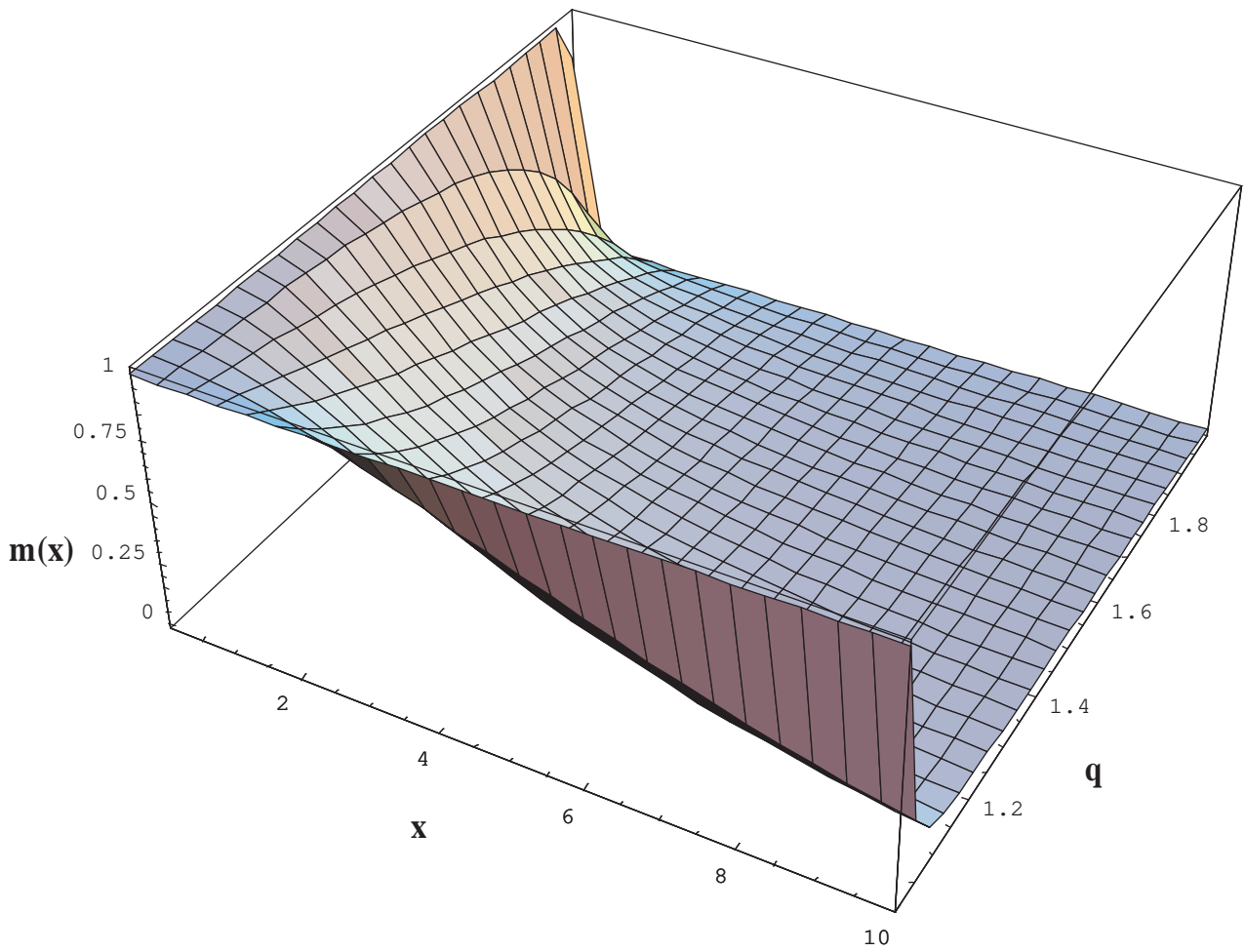
\end{document}